\documentclass[12pt]{article}
\usepackage{graphicx}
\usepackage{lineno}

\newcommand\pubnumber{SNSN-323-63}
\newcommand\pubdate{\today}

\def\institute{Stockholm University}

\def\Title#1{\begin{center} {\Large #1 } \end{center}}
\def\Author#1{\begin{center}{ \sc #1} \end{center}}
\def\Address#1{\begin{center}{ \it #1} \end{center}}

\newcommand\pubblock{\rightline{\begin{tabular}{l} \pubnumber\\
         \pubdate  \end{tabular}}}
\newenvironment{Abstract}{\begin{quotation}  }{\end{quotation}}
\newenvironment{Presented}{\begin{quotation} \begin{center} 
             PRESENTED AT\end{center}\bigskip 
      \begin{center}\begin{large}}{\end{large}\end{center} \end{quotation}}
\def\Acknowledgements{\bigskip  \bigskip \begin{center} \begin{large}
             \bf ACKNOWLEDGEMENTS \end{large}\end{center}}

\begin{document}
\begin{titlepage}
\pubblock

\vfill
\Title{Modelling $Wt$ and $tWZ$ production at NLO for ATLAS analyses}
\vfill
\Author{ Olga Bessidskaia Bylund \\ on behalf of the ATLAS Collaboration}
\Address{\institute}
\vfill
\begin{Abstract}
The generation of single top production in the $Wt$ channel at NLO gives rise to doubly-resonant diagrams that overlap with $t \bar t$, and similarly for associated production with a $Z$ or Higgs boson. Several methods exist to remove this overlap, thus avoiding double counting and separating the two processes.
In Diagram Removal 1, the amplitudes of doubly-resonant diagrams are set to zero, which also removes the interference of $Wt$ with $t \bar t$. If a measurement is performed in a region where this interference is not negligible, it would be of interest to take it into account. Another method, Diagram Subtraction, preserves the interference. However, it relies upon momentum reshuffling, which introduces uncertainties to the prediction. An alternative method, Diagram Removal 2, does not involve reshuffling and preserves the interference by subtracting the amplitude squared of the doubly-resonant diagrams from the total.
The predictions of Diagram Removal 2 from MG5\_aMC@NLO are compared with other predictions for $Wt$ and $tWZ$ production in the context of analyses at the ATLAS experiment. Implications of using Diagram Removal 2 are discussed, both for measurements of single top processes and when $Wt$ or $tWZ$ are treated as a background. Differential distributions are shown, investigating in what regions the predictions differ the most.
\end{Abstract}
\vfill
\begin{Presented}
$9^{th}$ International Workshop on Top Quark Physics\\
Olomouc, Czech Republic,  September 19--23, 2016
\end{Presented}
\vfill
\end{titlepage}
\def\thefootnote{\fnsymbol{footnote}}
\setcounter{footnote}{0}

\section{Introduction}
\begin{figure}[htb]
\centering
\includegraphics[height=3cm]{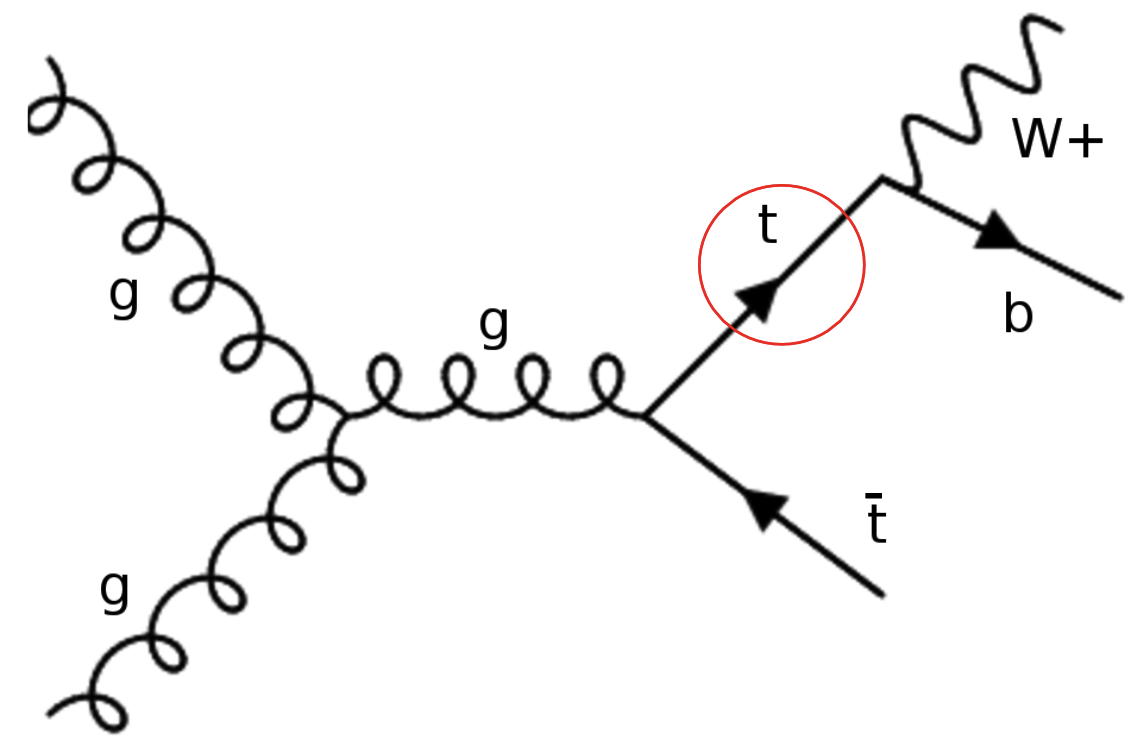} $\quad$
\includegraphics[height=3cm]{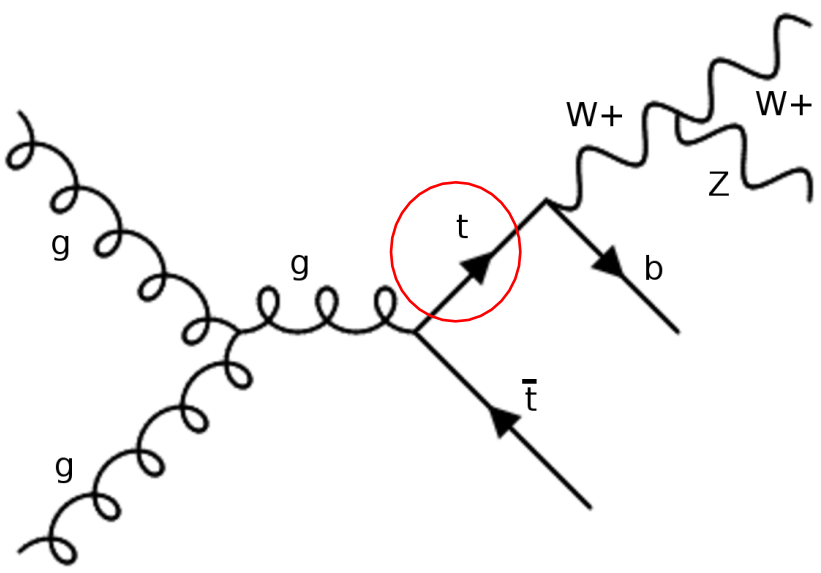} $\quad$
\includegraphics[height=3cm]{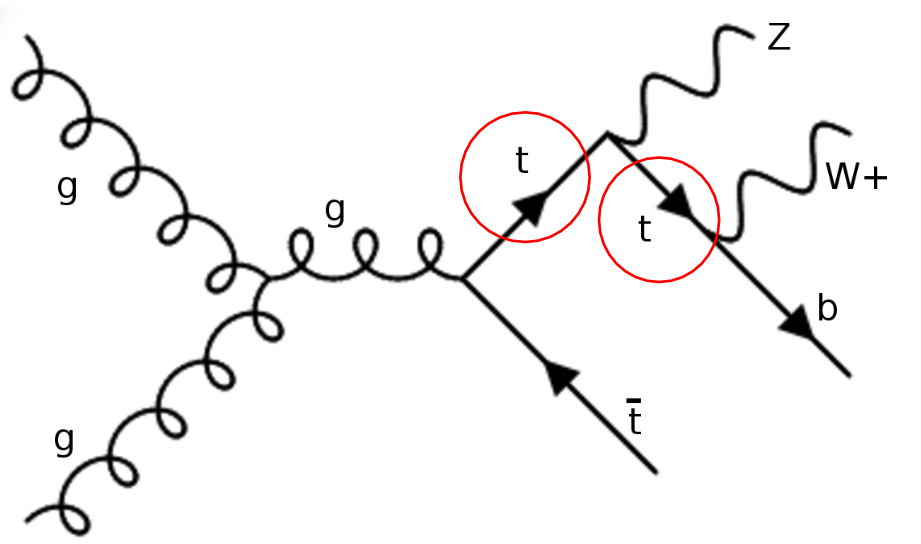}
\caption{Example Feynman diagrams for production of $Wt$ (left) and $tWZ$ (center, right) at NLO that overlap with $t \bar t$ or $t \bar t Z$. The presence of a top propagator, which gives rise to overlaps with $t \bar t$ or $t \bar t Z$, is indicated by a red circle.}
\label{fig:FD}
\end{figure}

The generation of single top production in the $Wt$ channel at NLO in QCD gives rise to Feynman diagrams that overlap with those appearing in $t \bar t$ production. Diagrams with a top quark propagator connected to a $W$ boson and a $b$ quark, as shown in Fig. \ref{fig:FD} (left), would also be generated in on-shell production of $t \bar t$, followed by top quark decay. In order to separate $Wt$ and $t \bar t$ and avoid double counting, the overlap between the processes needs to be addressed.

With what is commonly known as Diagram Removal, here referred to as DR1, the amplitudes of the overlapping diagrams are set to zero. This removes the overlap, as well as the interference of $Wt$ with $t \bar t$. The matrix element for $Wt$ is written in terms of singly-resonant (sr) and overlapping doubly-resonant (dr) contributions in Eq. (\ref{eq:dr1}). Removing the overlap at the level of the matrix element retains only the first term in Eq. (\ref{eq:dr2}):
\vspace{-0.5cm}
\begin{eqnarray}
\mathcal M_{tot} = \mathcal M_{sr} + \mathcal M_{dr},  \label{eq:dr1} \\
\mathrm{DR1} \qquad \qquad \qquad \qquad \qquad \qquad \nonumber \\
\mathcal |\mathcal M_{tot}|^2 = \underbrace{\overbrace{|\mathcal  M_{sr} |^2}+ 2 Re(\mathcal M^*_{sr}\mathcal M_{dr})} + |\mathcal M_{dr}|^2 \label{eq:dr2} \, .\\
\mathrm{DR2} \qquad \qquad \qquad \qquad  \nonumber
\end{eqnarray}
Another method, called Diagram Subtraction (DS) \cite{DS1}, removes the overlap at the level of the cross section, while including the interference in the prediction. This method is gauge invariant by construction, but relies upon momentum reshuffling, which introduces uncertaines to the prediction. An alternative method is Diagram Removal 2 (DR2) \cite{tWH}, where the matrix element squared of the overlapping diagrams is subtracted from the square of the complete expression, thus keeping the interference without introducing momentum reshuffling, as illustrated in Eq. (\ref{eq:dr2}).

In Ref. \cite{tWH}, it is shown that applying DR2 to $Wt$ and $tWH$ in MG5\_aMC@NLO \cite{Madgraph} yields results that are consistent with gauge invariant predictions that assess the interference. Thus one can generate $Wt$ and $tWH$ events at NLO, including the interference, without relying on momentum reshuffling. In the work described here, the DR2 method is extended to the $tWZ$ porcess and applied to Monte Carlo samples with kinematic cuts and selections applied to approach the signal regions used for analyses by the ATLAS Collaboration \cite{atlas}; a fuller description is given in Ref. \cite{conf}.

For the $tWZ$ process, several different kinds of doubly-resonant diagrams contribute, as shown in Fig. \ref{fig:FD} (center, right). In addition to diagrams that resemble $t \bar t Z$ production, followed by a top quark decay $t \rightarrow W b$, there are diagrams mimicking production of $t \bar t$ followed by a three body decay $t \rightarrow Z W b$ and diagrams that can be viewed as both of these types. That is, $tWZ$, $t \bar t Z$ and $t \bar t$ all interfere with each other. Due to the different contributions to the interference, it is challenging to verify that DR2 properly describes the effects.
\vspace{-0.4cm}
\section{Results and discussion}

\begin{figure}[htb]
\centering
\includegraphics[height=5cm]{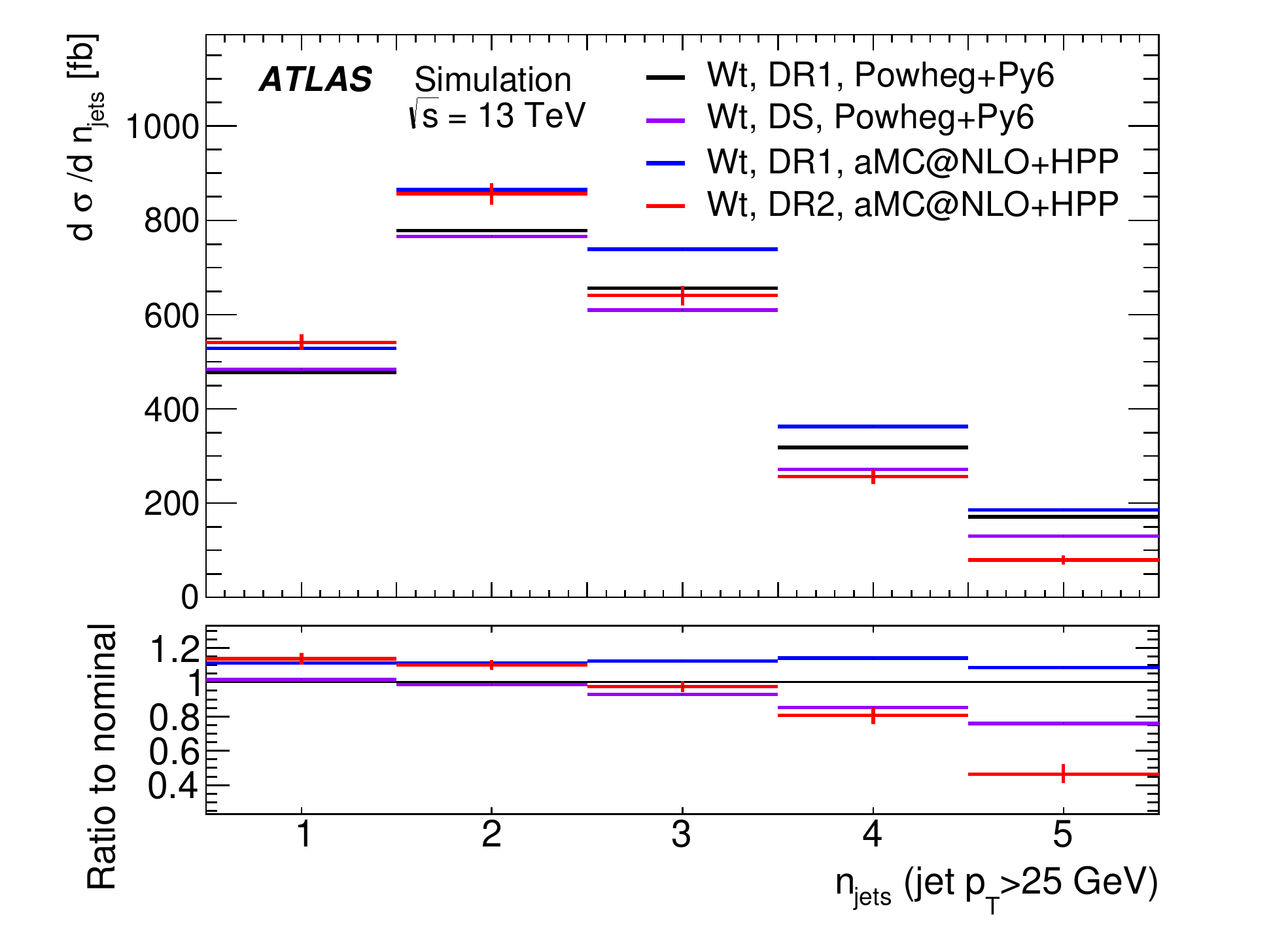}
\includegraphics[height=5cm]{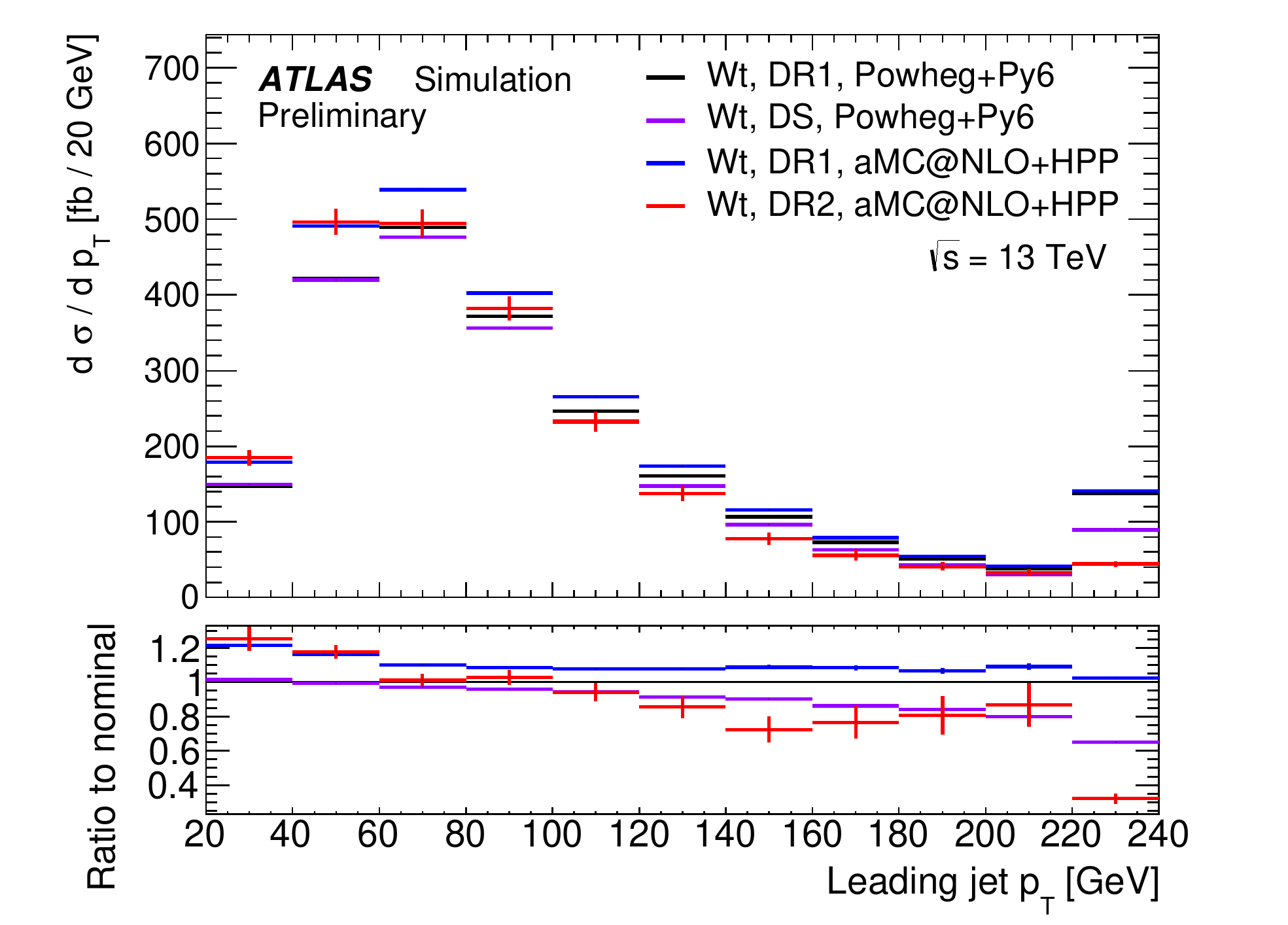}
\caption{Jet multiplicity (left) and leading jet $p_T$ (right) for $Wt$ \cite{conf}. The nominal DR1 sample in Powheg is shown in black, the DS Powheg sample in purple, the DR1 MG5\_aMC@NLO sample in blue and the DR2  MG5\_aMC@NLO prediction in red. The overflow is included in the last bin.}
\label{fig:wt_jet}
\end{figure}

The DR1 and DR2 predictions are compared, using MG5\_aMC@NLO for the event generation, with the top quarks and $W$ bosons required to decay leptonically in MadSpin and parton showering is provided by Herwig++. For comparison, the nominal ATLAS sample is also shown, which uses \textsc{Powheg-box} \cite{Powheg} with DR1 and interfaced with Pythia6 shower, as well as a similar sample that instead applies DS. 
From Table \ref{tab:wt}, the interference of $Wt$ with $t \bar t$ is destructive and found to be of the order of 10 \%, in agreement with Ref. \cite{tWH}.
\begin{table}[h!]
\begin{center}
\scalebox{0.9}{
\begin{tabular}{|c|c|c|}
\hline
Method & Total cross section & Difference w.r.t. DR1\\
\hline
MG5\_aMC@NLO DR1 & 7.87 pb & \\
MG5\_aMC@NLO DR2 & 7.09 pb & -10 \% \\ 
\hline
\textsc{Powheg-Box} DR1 & 7.16 pb &  \\
\textsc{Powheg-Box} DS & 6.82 pb & -5 \% \\
\hline
\end{tabular}
}
\caption{Inclusive cross section for $Wt$. Four predictions are compared: DR1 and DS evaluated with Powheg and DR1 and DR2 with MG5\_aMC@NLO. The difference with respect to the DR1 predictions for for each generator is shown.}
\label{tab:wt}
\end{center}
\end{table}
\vspace{-0.5cm}
The selection for the distributions shown in Fig. \ref{fig:wt_jet} requires two leptons with $p_T>20$ GeV, $|\eta|<2.5$ and the jets have transverse momentum $p_T>25$ GeV and pseudorapidity $|\eta|<2.5$. These selections are designed to be similar to the preselection for the $Wt$ measurement \cite{wt_m} by ATLAS.
The predictions that evaluate the interference (DR2, DS) decrease with jet multiplicity compared to the predictions that do not (DR1). The difference at high jet multiplicity is consistent with the explanation that diagrams that overlap with $t \bar t$ always produce a $b$ quark in addition to the $W$ boson and top quark, while singly-resonant diagrams can be produced without an additional quark. Moreover, the interference is larger in the presence of a jet with high $p_T$, which can be seen in the leading jet $p_T$ spectrum.

The event generation for the $tWZ$ process is performed with MG5\_aMC@NLO, interfaced with Madspin for particle decays, with the $Z$ boson required to decay into two charged leptons. Parton shower is provided by Pythia8. 
The difference in the total prediction with DR1 and DR2 is considerably larger than for $Wt$: 28 \% inclusively, see Table \ref{tab:twz}. This is of a similar magnitude to the interference of $tWH$ with $t \bar t H$ \cite{tWH}.
\begin{table}[h!]
\begin{center}
\scalebox{0.9}{
\begin{tabular}{|c|c|c|}
\hline
Method & Cross section & Difference w.r.t. DR1\\
\hline
MG5\_aMC@NLO DR1  & 15.6 fb & \\
MG5\_aMC@NLO DR2 & 12.2 fb & -28\% \\ 
\hline
\end{tabular}
}
\caption{The total cross section for the two $tWZ$ samples, with the $Z$ boson required to decay to two charged leptons. The difference with respect to the DR1 prediction is shown.}
\label{tab:twz}
\end{center}
\end{table}
\vspace{-0.5cm}

\begin{figure}[htb]
\centering
\includegraphics[height=5cm]{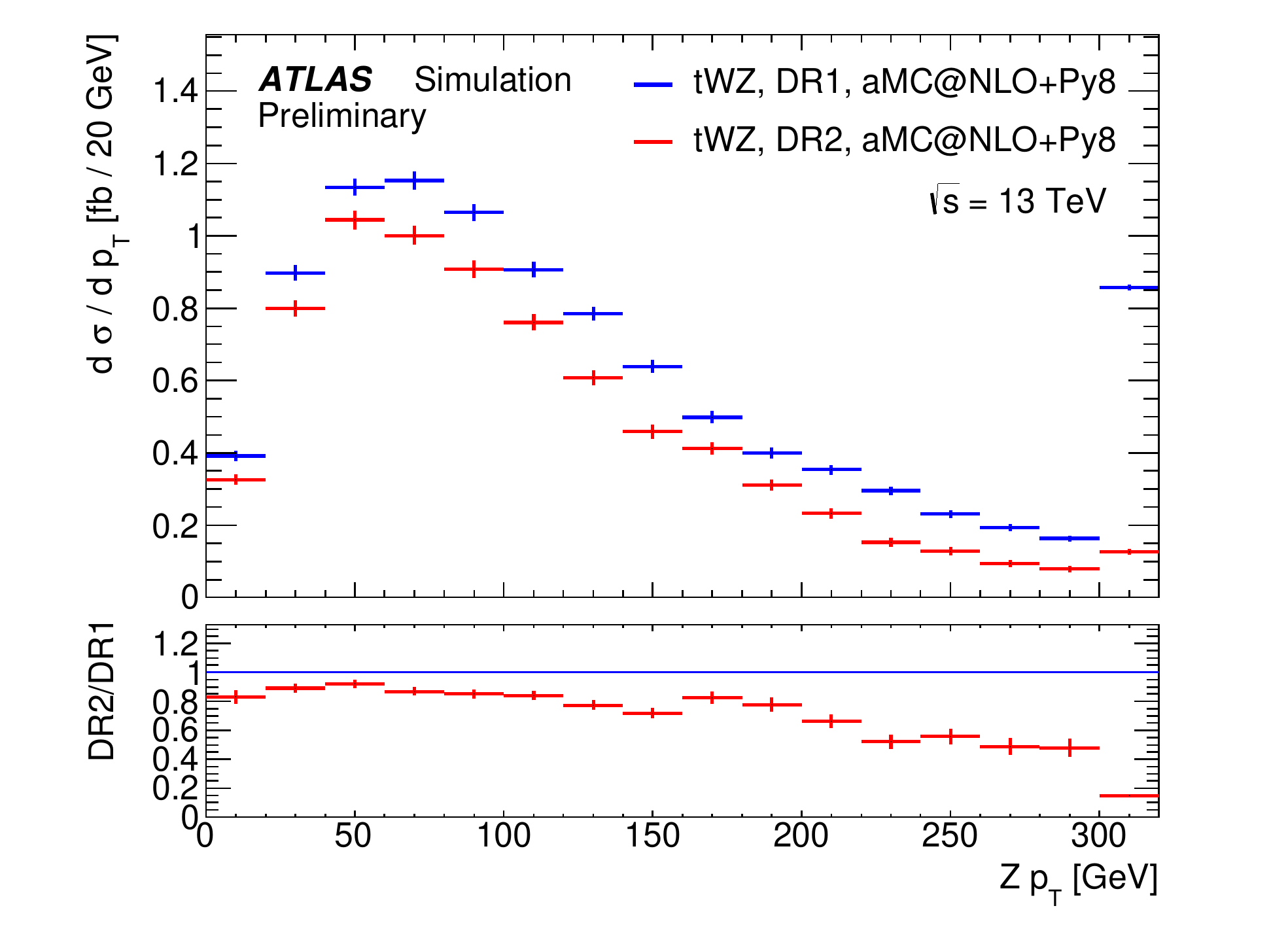}
\includegraphics[height=5cm]{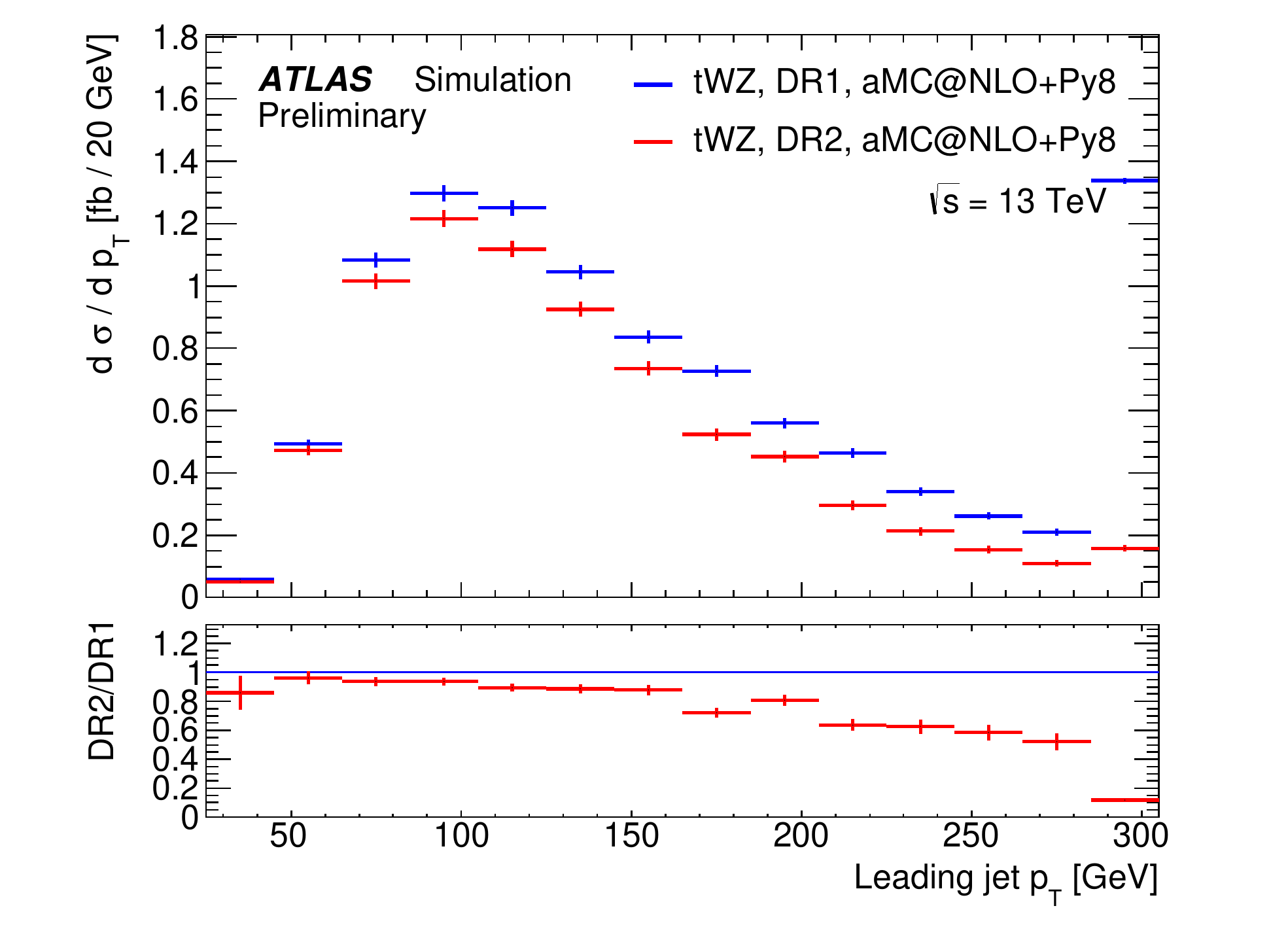}
\caption{The $p_T$ of the $Z$ boson (left) and the $p_T$ of the leading jet (right) for $tWZ$ \cite{conf}. Both predictions are produced using MG5\_aMC@NLO. The DR1 prediction is shown in blue and the DR2 prediction is shown in red. The overflow is included in the last bin.}
\label{fig:twz_pt}
\end{figure}

In Fig. \ref{fig:twz_pt}, a selection is applied, requiring at least two charged leptons with $p_T>15$ GeV and $|\eta|<2.5$ and at least three jets with $p_T > 25$ GeV, $|\eta|<2.5$. This selection is chosen to be close to the selection for the $t \bar t Z$ analyses by ATLAS \cite{ttz_1}, where $tWZ$ appears as a major background. At high $p_T$ of the $Z$ boson, the difference in prediction increases as well as at high $p_T$ for the leading jet.

In conclusion, the effect of Diagram Removal 2 has been studied on $Wt$ and $tWZ$ production in regions near the selections of ATLAS analyses. Differences are found both inclusively and differentially, which should be taken into account when modelling these processes at NLO and assessing the subsequent theoretical uncertainties.

\vspace{-5mm}

\Acknowledgements
Many thanks to Federico Demartin for demonstrating the Diagram Removal 2 method and for many discussions.
I am deeply grateful to the K \& A Wallenbergs Jubileumsfond that enabled me to attend the Top Workshop.


\begin{thebibliography}{99}






\bibitem{DS1}
S. Frixione et al., Single-top hadroproduction in association with a W boson, JHEP 0807 (2008) 029.

\bibitem{tWH}
F. Demartin et al., tWH associated production at the LHC, 2016, arXiv: 1607.05862.

\bibitem{atlas}
ATLAS Collaboration, The ATLAS Experiment at the CERN Large Hadron Collider, JINST 3 (2008)
S08003.

\bibitem{conf}
The ATLAS Collaboration, Studies on top-quark Monte Carlo modelling for Top2016, 2016, ATL-PHYS-PUB-2016-020.

\bibitem{Madgraph}
J. Alwall et al., The automated computation of tree-level and next-to-leading order differential cross sections, and their matching to parton shower simulations, JHEP 1407 (2014) 079.


\bibitem{Powheg}
S. Frixione et al., A positive-weight next-to-leading-order Monte Carlo for heavy flavour hadro- production, JHEP 0709 (2007) 126

\bibitem{wt_m}
ATLAS Collaboration, Measurement of the cross-section of the production of a W boson in association with a single top quark with ATLAS at $\sqrt s$ = 13 TeV, ATLAS-CONF-2016-065, 2016.

\bibitem{ttz_1}
ATLAS Collaboration, Measurement of the $t \bar t Z$ and $t \bar t  W$ production cross sections in multilepton final states using 3.2 $\mathrm{fb}^{-1}$ of pp collisions at $\sqrt s$ = 13 TeV with the ATLAS detector, 2016, arXiv: 1609.01599 [hep-ex].


\end{thebibliography}
\end{document}